# Cortical recruitment and functional dynamics in postural control adaptation and habituation during vibratory proprioceptive stimulation

KJ Edmunds[1], H Petersen[2,3], M Hassan[4], S Yassine[4], A Olivieri[5], F Barollo[1,6], R Friðriksdóttir[1], P Edmunds[1], MK Gíslason[1], A Fratini[6], and P Gargiulo[1,7]

[1] Institute for Biomedical and Neural Engineering, Reykjavík University, Reykjavík, Iceland
[2] Department of Anatomy, University of Iceland, Reykjavík, Iceland
[3] Akureyri Hospital, Akureyri, Iceland
[4] University of Rennes, LTSI - U1099, F-35000 Rennes, France
[5] Dept. of Electrical Eng. and Information Tech., University of Naples Federico II, Naples, Italy
[6] School of Life and Health Sciences, Aston University, Birmingham, UK
[7] Department of Science, Landspítali, Reykjavík, Iceland

7/9/2018
**Address for Correspondence:**

*Dr. Kyle J. Edmunds*
Institute for Biomedical and Neural Engineering, Reykjavík University
Menntavegur 1, 101 Reykjavík, Iceland
+354 788 9320
Email: kyle14@ru.is

## -Abstract

***Objective.*** Maintaining upright posture is a complex task governed by the integration of afferent sensorimotor and visual information with compensatory neuromuscular reactions. The objective of the present work was to characterize the visual dependency and functional dynamics of cortical activation during postural control.

***Approach.*** Proprioceptic vibratory stimulation of calf muscles at 85 Hz was performed to evoke postural perturbation in open-eye (OE) and closed-eye (CE) experimental trials, with pseudorandom binary stimulation phases divided into four segments of 16 stimuli. 64-channel EEG was recorded at 512 Hz, with perturbation epochs defined using bipolar electrodes placed proximal to each vibrator. Power spectra variation and linearity analysis was performed via fast Fourier transformation into six frequency bands ($\Delta$, 0.5-3.5 Hz; $\theta$, 3.5-7.5 Hz; $\alpha$, 7.5-12.5 Hz; $\beta$, 12.5-30 Hz; $\gamma\_low$, 30-50 Hz; and $\gamma\_high$, 50-80 Hz). Finally, functional connectivity assessment was explored via network segregation and integration analyses.

***Main Results.*** Spectra variation showed waveform and vision-dependent activation within cortical regions specific to both postural adaptation and habituation. Generalized spectral variation yielded significant shifts from low to high frequencies in CE adaptation trials, with overall activity suppressed in habituation; OE trials showed the opposite phenomenon, with both adaptation and habituation yielding increases in spectral power. Finally, our analysis of functional dynamics reveals novel cortical networks implicated in postural control using EEG source-space brain networks. In particular, our reported significant increase in local $\theta$ connectivity may signify the planning of corrective steps and/or the analysis of falling consequences, while $\alpha$ band network integration results reflect an inhibition of error detection within the cingulate cortex, likely due to habituation.

***Significance.*** Our findings principally suggest that specific cortical waveforms are dependent upon the availability of visual feedback, and we furthermore present the first evidence that local and global brain networks undergo characteristic modification during postural control.

# 1. Introduction

Maintaining upright posture is a complex task governed by the integration of afferent sensorimotor information with compensatory neuromuscular reactions [1], and the compensation for unpredictable perturbations in balance is essential to retaining stability and avoiding injury from falling. The cerebral cortex and central nervous system (CNS) play integral roles in postural control, incorporating information from visual, somatosensory, and vestibular systems to carry out the corrective motions needed to maintain balance [2-4]. The role of the CNS and subcortical structures in this regard is well-documented in literature for the innate generation of feedforward and feedback adaptive adjustments to reduce the risk of balance loss [5-7]. However, understanding the role of the cerebral cortex has been of comparatively recent focus, along with its potential relationship with visual subsystems. Nonetheless, there is growing evidence for the importance of cortical involvement in adapting to transient balance perturbation [4-6, 8-13].

## 1.1. Role of the Cerebral Cortex in the Postural Control System

The contributions from infratentorial or supratentorial structures in maintaining upright posture has been of much recent focus in vestibular research involving human postural biomechanics. In such studies, postural control feedback is generalized under single inverted pendulum (SIP) theory, wherein the simplest spinal model (containing one or few spinal segments) is subjected to reflex controls aimed at maintaining a constant body center of pressure (COP) on a limited base of support [14-16]. This SIP model has been shown to accurately represent postural control during both small body perturbations inherent to bipedal quiet stance [17], as well as movement when the body is sufficiently far from its limits of antero-posterior or medial-lateral stability [18]. In general, when upright stance is perturbed, infratentorial reflexes trigger synchronized, efferent muscle reactions within the spine; these reflexes are commensurately registered by supratentorial (cortical) structures. This combination of feedback systems during postural perturbation is a complex relationship that further involves central and peripheral nervous structures [20]; as such, optimizing the validity of extant postural control models presents a strategic priority for research [19-21].

Characterizing the precise role of the cerebral cortex in governing the supratentorial response to postural perturbation has been addressed by research on sensorimotor integration, which has specifically implicated the posterior-parietal cortex for its interconnectivity with motor and premotor cortices [2, 22], and further studies have suggested the region's role in complex sensorimotor processing [23-25]. In addition, enhanced activation of the fronto-central region has been reported during the visual recognition of one's postural instability, both with and without warning cues [4, 5, 26], as well as during self-initiated postural changes [6, 13, 27]. In particular, the anterior cingulate cortex (ACC) in the fronto-central region is well-known for its crucial involvement in action monitoring and the detection of error signals, specifically increasing in activity during the detection of balance instability [28-30].

While many tools have been employed to study neurocognitive processes in this regard, the use of electroencephalography (EEG) as a neuroimaging technique remains disparately reported. Nonetheless, extant literature has investigated the utility of EEG through examination of evoked potentials associated to balance perturbation. Most typically, changes in evoked activity have been reported as time-domain event-related potentials (ERP) and/or perturbation-evoked responses (PER), and ERP/PER components such as N1 amplitudes and Contingent Negative Variation (CNV) have been reported [4-6, 12, 13]. While ERP responses have exhibited a wide distribution during unpredictable perturbations to upright posture, the propensity of literature suggests the generalizability of responses to fronto-central and occipital sites [59-61]. Power spectral density (PSD) analysis represents an alternative, yet analogously conventional method for EEG signal analysis, which remains comparatively underreported in postural control literature, despite its proven utility in cognitive and/or motor task EEG studies [28, 31-33].

In addition to ERP analysis, emerging evidence highlights the utility of EEG for the estimation of functional brain network dynamics; indeed, functional connectivity in this regard plays a crucial role in many cognitive and motor functions [34]. In addition to its non-invasiveness and the relative ease of use (portability), a key advantage of EEG in discerning functional neural connectivity is its excellent temporal resolution, which offers the unique opportunity to both track large-scale brain networks over very short durations and to analyze fast dynamic changes that can occur during resting states or in brain disorders [35].

## 1.2. Comparative Indicators for the Cortical Response: Adaptation and Habituation

There are two primary indicators for cortical response in balance perturbation and postural control: cortical *adaptation* and *habituation*. Cortical adaptation in postural control may be described as transient changes in motor control strategy when exposed to acute changes in balance or sensorimotor input [36]. Cortical habituation, conversely, may be defined as the process of gradual improvement in postural rescue acquired through repetitive adaptation, through which enhanced motor control is acquired from repetitive, strategic corrections [37, 38]. Indeed, the contributory mechanisms of vestibular and subcortical systems represent phylogenically older structures for governing balance and employing learned motor movements, altogether suggesting their comparatively larger roles in postural adaptation [39]. In general, postural control is transiently and longitudinally improved by repetitive exposure to postural perturbation, such as in daily exercise, which encourages both functional and structural adaptation in the neuromuscular system [31, 40, 41]. In a research context, external vibratory stimulation applied to a muscle or tendon is a typical technique for the artificial induction of postural change, evoking the activation of proximal muscles to instigate an erroneous sensation of movement, which ultimately induces a commensurate postural response [33, 42].

Maintaining vertical posture may be generalized by the complex integration of feed-forward and feed-back mechanisms in cortical and subcortical structures to generate forces toward a supporting surface; this is typically achieved by calculated adjustments to the position of body segments or extremities. However, the precise role of the cerebral cortex and functional dynamics during the adaptation or influence of motor neuron executions remains debated; to our knowledge, there exists no study on the assessment of changes in EEG power spectra and functional connectivity during proprioceptic stimulation. Furthermore, while vision has been known to play an important role in postural control, its relationship with cortical activity during postural perturbation is not well understood. These notions altogether motivate the present work, wherein we report the employment of EEG to investigate the cortical dynamics of postural adaptation and habituation with and without visual feedback following vibratory proprioceptive stimulation of calf muscles.

## 2. Material and Methods

### 2.1. Experiment Setup

The main experiment was performed in the Icelandic Center for Neurophysiology at Reykjavik University. Ten healthy subjects between 22 and 25 years old were instructed to maintain upright posture during vibratory balance perturbations evoked by vibrators designed to deliver simultaneous stimulation to both calf muscles. These vibrators comprised of revolving DC-motors equipped with a 3.5 g eccentric weight, contained in cylindrical casing measuring 0.06 m length and 0.01 m in diameter. Each vibrator was held in place by elastic straps fastened tightly around the widest point of the calf muscle, and electrical stimulation was set to deliver a vibratory amplitude of 1.0 mm and a vibration frequency of 85 Hz. Stimulations were applied according to a pseudorandom binary sequence schedule, where each shift had a random duration of 0.8 seconds to 6.4 seconds, yielding an effective stimulus bandwidth of 0.1-2.5 Hz.

The subjects stood on a pressure platform where postural sway was monitored by recording anterio-posterior and medial-lateral stance pressure changes [43]; it is important to note that although sway and its corresponding EEG may be the product of the same perturbance, these responses are not necessarily dependent upon each other. As such, postural sway was monitored here as a quality control method to ensure appropriate stimulation response. Simultaneous to postural sway, EEG data were recorded continuously from 64 channels in bipolar configuration, with each experimental trial segmented into two phases lasting a total of 230 seconds: the quiet stance phase (30 seconds), designated as the prestimulus baseline ($BL$), and the stimulation phase (200 seconds). Each stimulation phase was subdivided into four segments, $Q1$-$Q2$-$Q3$-$Q4$, each containing 16 stimuli (64 in-total per experimental trial). In particular for this study, $Q1$ and $Q4$ segments were separately isolated to define adaptation and habituation epoch events, respectively. Finally, experimental trials were performed twice for each subject according to two conditions: with closed eyes (CE) and with open eyes (OE). Experimental trial order was randomized for each subject in the cohort to reduce the effect of learning due to repeated stimulus, as described [58].

## 2.2. EEG Data Acquisition

The acquisition of EEG data was performed using a 64-channel wet-electrode cap in bipolar configuration with a portable amplifier and recording tablet using EEGO software (ANT Neuro, Enschede Netherlands). A standardized 10-20 system montage was used, and 3D head models were generated with each subject in accordance with standard montage electrode positions [44]. After strapping the calf muscle vibrators to both legs, participants were informed of the experiment protocol, and the EEG cap was attached, followed by the application of conductive electrode gel to reduce measured impedance to less than 10 kΩ in each channel. Next, a pair of bipolar electrodes were placed proximal to each vibrator to isolate and define vibratory epochs for each stimulation phase. After completing this instrumentation setup, the proprioceptive stimulation experiment was initialized, with EEG data acquired at a sampling frequency of 512 Hz.

## 2.3. Data Processing

Following data acquisition, recorded EEG was manually segmented for subject into OE and CE recordings. Next, each dataset was processed according to a series of transformation operations to prepare the measurements for analysis in a customized GUI in Matlab (MathWorks, Inc., Natick, Massachusetts, USA). Initial preprocessing operations were performed as outlined:

1. **Baseline and Stimulation Phase Extraction**: each baseline ($BL$) and stimulation phase which were identified and extracted from each collected dataset to be separately processed.
2. **Vibratory Stimuli Detection**: using the signal from both bipolar calf electrodes, each of the 64 experimental vibrations were manually identified and divided into their four segments, $Q1$-$Q2$-$Q3$-$Q4$, as previously described. Subsequently, $Q1$ and $Q4$ segments were separately extracted from the signal in order to describe:
    a. *Adaptation*: the initial effects evoked by the proprioceptive stimulation ($BL$ vs. $Q1$).
    b. *Habituation*: the cortical conditioning process from prolonged postural perturbation ($Q1$ vs. $Q4$)

3. **Signal Filtering**: $BL$, $Q1$ and $Q4$ segments were filtered with a band-pass filter [0.3-80 Hz] with a 24dB/octave roll-off to isolate six EEG frequency bands ($\Delta$, 0.5-3.5 Hz; $\theta$, 3.5-7.5 Hz; $\alpha$, 7.5-12.5 Hz; $\beta$, 12.5-30 Hz; $\gamma_{low}$, 30-50 Hz; and $\gamma_{high}$, 50-80 Hz). Moreover, signals were filtered with a band stop filter from 49-51 Hz to remove undesired monomorphic artifacts from 50 Hz mains electricity.

4. **Artifact Detection**: automatic artefact detection using a manual voltage threshold set between -100μV and +100μV was used to exclude segments contaminated by eye blinks and/or motion artifacts. An artefact was detected automatically with the ASA feature "Artefact Detection", if EEG data exceed threshold values in at least one channel. The user interfaces of the FFT analysis features allow to "reject artefacts": this means that any events overlapping with artefacts was not used in the respective feature. If thresholded detection was inadequate for removing all visible artifacts, manual artefact identification was performed: all artefacts clearly contaminated by motion artefacts were marked and then excluded from Fourier analyses.

Finally, the following series of postprocessing operations were then executed:

1. **Fast Fourier Transformation (FFT):** FFT processing of each EEG dataset was performed using ASA software (ANT Neuro, Enschede Netherlands) using a frequency resolution of 0.977 Hz [62], and epoch durations set to 30 seconds for $BL$ data and the entire stimulus duration (as previously noted, ranging from 0.8-6.4 seconds) for $Q1$ and $Q4$ stimulation datasets.

2. **Power Spectrum Extraction**: power spectral density (PSD) values and topologies were then obtained from FFT analysis, normalized in ASA, and finally exported in ASCII format.

PSD values were exported in this regard for cohort comparison and statistical analysis using Matlab.

## 2.4. Power Spectra Variation

As previously stated, the primary objectives of the present investigation were to analyze cortical markers for OE and CE postural adaptation ($Q1 - BL$) and habituation ($Q4 - Q1$). To investigate these differences and clarify the role of the cerebral cortex in postural control, OE and CE difference spectra topologies were extracted for each EEG frequency band, showing electrodes and corresponding cortical regions with significant differences in utilization. In this comparison, significant changes in mean spectral power at each electrode were assessed using paired heteroscedastic t-tests, yielding topological $P$-value maps for each EEG waveform (with $P < 0.05$ the threshold for significance).

Furthermore, spectral power variations ($\delta\%$) were calculated for the cohort to assess whether percent changes in adaptation and habituation were significantly different depending on the availability of visual feedback. To do this, PSD values were extracted and averaged within each band over all electrodes, resulting in 6x1 vectors for each subject for both OE and CE trials, according to the following expressions:

**Adaptation**: $\delta(Q1 - BL)\% = 100 \times \frac{\mu_{Q1} - \mu_{BL}}{\mu_{BL}}$ (1)

**Habituation**: $\delta(Q4 - Q1)\% = 100 \times \frac{\mu_{Q4} - \mu_{Q1}}{\mu_{Q1}}$ (2)

To assess statistical significance between datasets, pairwise, two-sample Z-tests were performed, and $P < 0.05$ was analogously considered the threshold for significance. To correct for multiple comparisons, the $P$-values of sensor-wise comparisons were corrected using the Benjamini–Hochberg (BH) method [71].

### 2.5. Power Spectra Linearity

To assess the degree of correlation between PSD values in each recording epoch, mean spectral values for each member of the cohort were plotted against their own respective frequency band. For postural adaptation analysis, $BL$ and $Q1$ average values were compared for each frequency band, and for habituation assessment, $Q4$ and $Q1$ PSD values were compared. Pearson's correlation coefficients, $r$, were calculated for each correlation and were considered significantly correlated for $|r| > 0.7$ [63]. Furthermore, $P$-values were calculated using a homoscedastic, two-tailed Student's t-test of the null hypothesis that there exists no linear relationship, with $P < 0.05$ the threshold for significance.

### 2.6. Functional Connectivity Analysis

### 2.6.1. Brain Network Construction

Functional brain networks were constructed using the "EEG source connectivity" method, which is comprised of two main steps [45]: 1) reconstruction of cortical source dynamics by solving the inverse problem, and 2) estimation of the functional connectivity between reconstructed signals. Here, we used the weighted minimum norm estimate (wMNE) algorithm as the inverse solution [59]. The wMNE was computed, using Brainstorm [68] toolbox for a cortical mesh of 15000 vertices using openMEEG [69]. A Desikan-Killiany atlas-based segmentation approach was used, consisting in 68 cortical regions [70]. Time series within one region of interest were averaged after flipping the sign of sources with negative potentials. . Reconstructed regional time series were band-pass filtered across the six reported EEG frequency bands. Next, functional connectivity between regional time series was computed for each frequency band, using the phase locking value (PLV) metric, which returns a PLV value between 0 (no phase locking) and 1 (full synchronization) [46, 47].

### 2.6.2. Network Metrics

While functional connectivity provides key information about how different cortical regions are linked, graph theory analysis (GTA) offers a framework to characterize the network topology and organization. In practice, while many GTA metrics can be extracted from networks to characterize both local and global properties, we report the utility of two, in particular:

1. **Network Segregation (Local Information Processing)**: The clustering coefficient, $CC$, was computed as a direct measure of network segregation [64]. In brief, $CC$ denotes how close a node's neighbors tend to cluster together [65]. This coefficient is the proportion of connections among a node's neighbors, divided by the number of connections that could possibly exist between them, which is zero if no connections exist and one if all neighbors are connected.
2. **Network Integration (Global Information Processing)**: The relative importance of a node is proportional to its betweenness centrality, $C$, which reflects the proportion of the number of shortest paths in which the node participates (see [66] for a technical review).

All extracted network GTA measures were normalized with respect to random networks. Thus, we generated 500 surrogate random networks, derived from the original network, by randomly reshuffling the edge weights. Normalized values were computed by dividing original values by the average values computed on the randomized graphs [48]. Finally, the Jonckheere-Terpstra test was used to evaluate the trends of network GTA metrics as a function of the level of control (from $BL$ to $Q1$ through $Q4$). This non-parametric rank-based test can be used to determine the significance of a trend by the computation of a z-score equivalent assembled against its critical value (with $P < 0.05$ the threshold for significance). The statistical difference between conditions was performed at the level of each region (a node-wise analysis). Thus, the results are a set of brain regions that show significant differences between the conditions.

## 3. Results and Discussion

### 3.1. Power Spectra Variation and Linearity

Our first PSD analysis method involved the topological comparison of both average spectral power values for each EEG waveform, as well as *P*-values (*uncorrected*) to illustrate significant differences between spectral power in individual electrodes. **Figures 1-3** contain results from these analyses, showing comparisons between CE and OE conditions, in both adaptation ($Q1 - BL$) and habituation ($Q4 - Q1$).

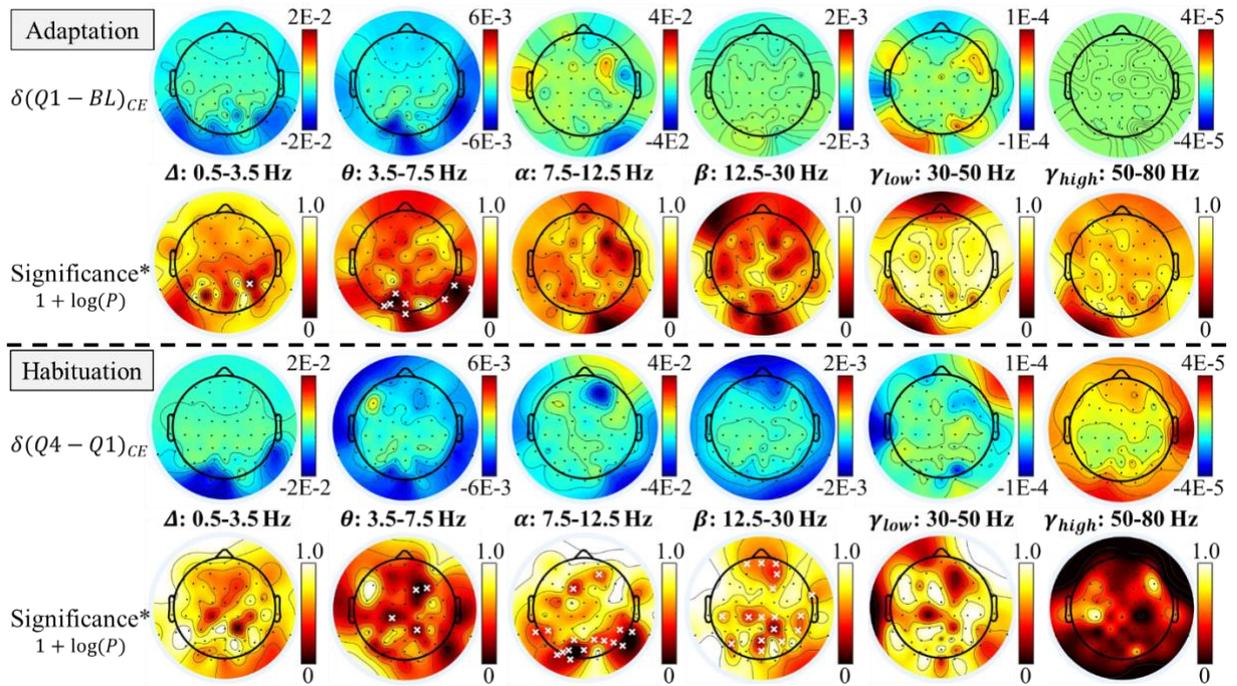

**Figure 1**: Spectral topology differences during CE postural adaptation ($Q1 - BL$) and habituation ($Q4 - Q1$). *Note statistical significance of $P < 0.05$ is denoted by each electrode denoted by a white 'x' in the significance maps.

As is evident in **Figure 1**, difference spectra values in CE adaptation varied considerably in specificity to each EEG waveform, with a general reduction in $Q1$ spectral power over $BL$ for lower frequencies ($\Delta$ and $\theta$), varied differences seen in $\alpha$ and $\beta$, and an increase in $Q1$ over $BL$ in $\gamma_{low}$ and $\gamma_{high}$. Significant spectral differences from reduced $Q1$ power were highlighted in occipital and right-temporal $\Delta$ and $\theta$. Contrastingly, higher waveforms yielded more positive differences in spectral power; while not significant, these differences may indicate a shift in cortical recruitment during CE postural adaptation from lower to upper EEG waveforms. In habituation trials, difference spectra illustrated a nearly global reduction in $Q4$ power over $Q1$, with the exception of $\gamma_{high}$ and right frontal $\gamma_{low}$. Significant spectral differences were found in $\theta$ around the precentral or superior frontal gyri, in addition to many locations across the cortex in $\alpha$ and $\beta$ bands; however, $\Delta$, $\gamma_{low}$, and $\gamma_{high}$ bands yielded minimal significance. These results show a generalized decrease in cortical activity in CE postural habituation, altogether suggesting the comparative employment of proprioceptive mechanisms over time.

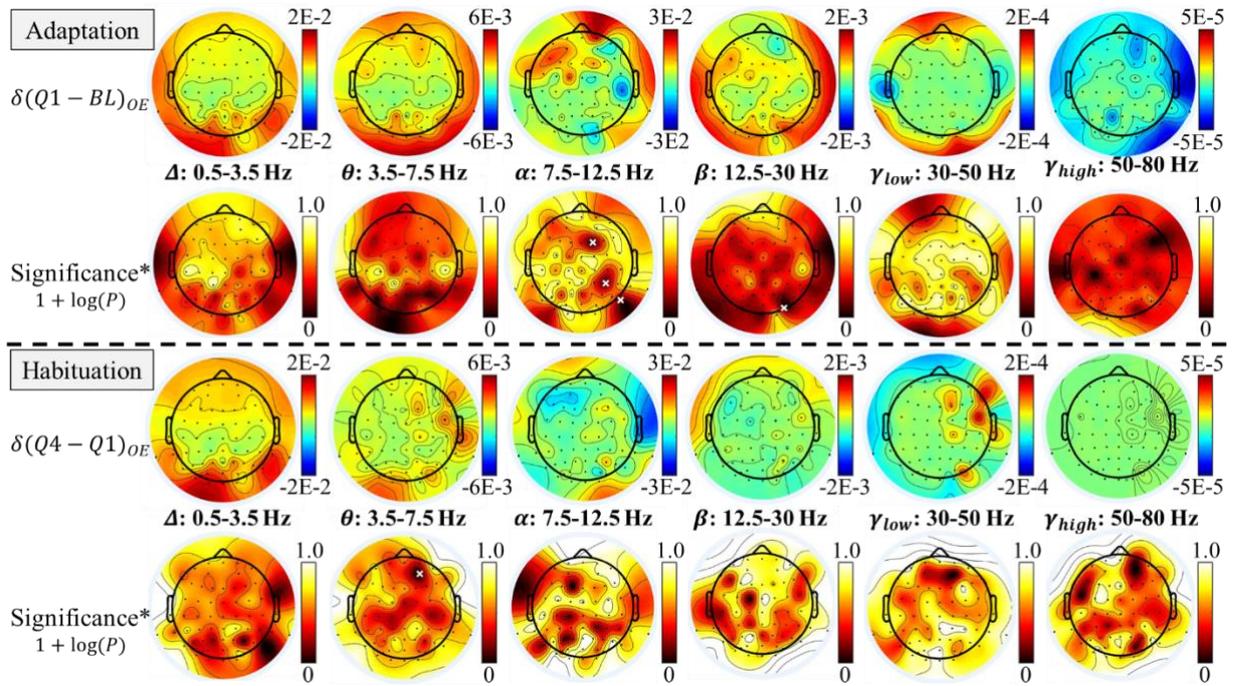

**Figure 2**: Spectral topology differences during OE postural adaptation ($Q1 - BL$) and habituation ($Q4 - Q1$). *Note statistical significance of $P < 0.05$ is denoted by each electrode denoted by a white 'x' in the significance maps.

As shown in **Figure 2**, OE adaptation difference spectra illustrated a nearly global increase in $Q1$ power over $BL$, with the exception of $\gamma_{high}$. These results are directly antithetical to our findings in CE adaptation, furthermore illustrating the inverse $\gamma_{high}$ power relationship observed in CE habituation. However, significant spectral differences (*uncorrected*) from increased $Q1$ power were limited to right-temporal and occipital $\alpha$ and $\beta$. Nonetheless, these results indicate that cortical recruitment may be directly related to the OE condition, as specifically evidenced by the generalized increase in $Q1$ activity over $BL$. In habituation trials, difference spectra again illustrated antithetical results to CE trials, showing an increase in $Q4$ power over $Q1$ in a majority of channels for each EEG band. However, $P$-value topologies indicated only one significant electrode in prefrontal $\theta$, suggesting the necessity for additional comparisons to solidify implicated differences between CE and OE trials (**Figure 3**).

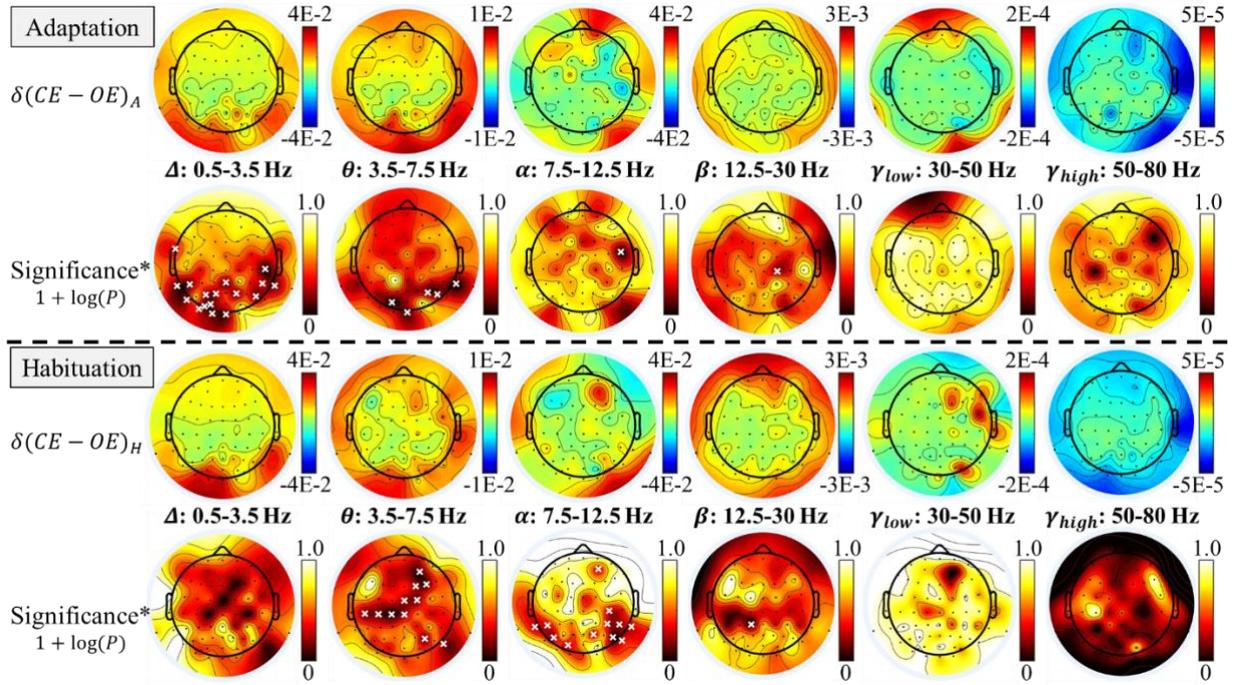

**Figure 3**: Spectral topology comparison between CE and OE difference spectra in both postural adaptation and habituation. *Note statistical significance of $P < 0.05$ is denoted by each electrode denoted by a white 'x' in the significance maps.

**Figure 3** depicts the comparison between OE and CE difference spectra, showing which trials incurred more cortical activity in both adaptation and habituation. Since a majority of CE trials resulted in negative spectral differences, while OE trials gave positive results (both in adaptation and habituation), it is logical to expect a majority of these $\delta(OE - CE)$ values to be positive; indeed, this is precisely what we observe. These differences are statistically significant in occipital and temporal electrodes for $\Delta$ and $\theta$ bands during adaptation, with additional significant electrodes at the precentral gyrus in $\alpha$ and $\beta$ bands. Further significance during habituation is evident in $\theta$ at the left-precentral gyrus, as well as the prefrontal and right-temporal cortices. Significance in habituation spectra was also found in the $\alpha$ band through both occipital and right-posterior parietal electrodes. Finally, while again showing fewer significant electrodes, habituation $\beta$ differences yielded significance in one central parietal electrode, with no significance in higher frequency waveforms. These results altogether show that the disparate trends observed in OE and CE difference spectra are statistically significant, suggesting the observation that cortical activity during postural control may be significantly higher with the use of vision. These findings altogether directly support our hypothesis that visual recognition of instability plays a critical role in governing the employment of cortical processes for postural control.

Following topological comparisons, pairwise spectral power variation ($\delta\%$) was calculated to assess whether percent changes in adaptation and habituation were significantly different, depending on the availability of visual feedback. The following tables contain results from this analysis, detailing the pairwise, two-sample Z-tests with significance designated for $P < 0.05$ and $P < 0.01$ between CE and OE adaptation and habituation trials.

**Table 1.** PSD pairwise variation in open eyes (OE) and closed eyes (CE) postural adaptation ($Q1 - BL$) and habituation ($Q4 - Q1$) spectral comparisons

| | Frequency Band Pairwise Variation in Mean PSD[1] | | | | | |
|---|---|---|---|---|---|---|
| **Adaptation:** | $\Delta$ | $\theta$ | $\alpha$ | $\beta$ | $\gamma_{low}$ | $\gamma_{high}$ |
| $\delta(Q1 - BL)_{OE}$ | 83,5% | 46,8% | 13,0% | 48,6% | 11,9% | -93,1%[†] |
| $\delta(Q1 - BL)_{CE}$ | -6,5%[†] | -30,9%[†] | -3,5% | -0,9%[†] | -8,2% | 28,0%[†] |
| **z-score** | 3,461** | 0,729 | 0,772 | 2,472* | 0,275 | 2,978** |
| **Habituation:** | $\Delta$ | $\theta$ | $\alpha$ | $\beta$ | $\gamma_{low}$ | $\gamma_{high}$ |
| $\delta(Q4 - Q1)_{OE}$ | 53,4%[†] | 24,7%[†] | -2,3% | 1,2% | 12,8%[††] | 35,2%[††] |
| $\delta(Q4 - Q1)_{CE}$ | -36,6% | -31,6% | -20,8% | -43,9% | -3,7% | -13,4% |
| **z-score** | 0,755 | -0,343 | -1,294 | -2,285* | 0,740 | 1,137 |

[1]Waveform bandwidths: $\Delta$, 0.5-3.5 Hz; $\theta$, 3.5-7.5 Hz; $\alpha$, 7.5-12.5 Hz; $\beta$, 12.5-30 Hz; $\gamma_{low}$, 30-50 Hz; and $\gamma_{high}$, 50-80 Hz
z-test: *$P < 0.05$; **$P < 0.01$
Pearson's coefficient (linearity): [†]$R > 0.7$; [††]$R > 0.9$

The results presented in **Table 1** carry several important EEG waveform distinctions. Firstly, both OE adaptation and habituation tests indicated an increase in spectral power across five of the six bands (exceptions being $\gamma_{high}$ and $\alpha$, respectively), while CE tests resulted in decreases in power over all habituation bands, along with all adaptation bands, except $\gamma_{low}$. These findings are in direct accordance with the results shown in **Figures 1-3**, suggesting again that, following both acute and prolonged proprioceptive perturbation, cortical activity is upregulated with the availability of visual feedback, while conversely downregulated without vision. These notions further support our hypothesis that visual recognition of instability plays a critical role in governing the employment of cortical processes for postural control.

It is likewise intriguing to note both the significant difference between $\gamma_{low}$ and $\gamma_{high}$ values and the degree of independence between $\gamma_{low}$ and $\gamma_{high}$ PSD values, overall. Extant literature has suggested that generalized $\gamma$ activity may precede the initiation of compensatory backward postural movement when balance is in danger [29], but our results extend this notion with the segmentation of our $\gamma$ bands; it is evident here that the cortical response to instability with visual feedback engaged the $\gamma_{low}$ portion of the $\gamma$ waveform, whereas the $\gamma_{high}$ waveform was almost completely downregulated in adaptation, yet upregulated with prolonged perturbation. Furthermore, antithetical to other bands, $\gamma_{high}$ spectral power increased significantly ($P < 0.01$) in CE adaptation trials, suggesting its specificity as a cortical activity marker when vision is unavailable. However, it is important to likewise note that, although these results present clear cortical differences between $Q1$ and $Q4$ which have are referenced under the adaptation and habituation paradigm in literature [4], the onset of habituation still remains differentially possible throughout the course of stimulation phases – a notion which motivates further research in this regard.

In addition to the $\gamma$ bands, the reported results likewise show the significant ($P < 0.05$) drop in $\beta$ waveform spectral power in both adaptation and habituation, but with strikingly disparate characteristics. In adaptation, OE trials yielded a large upregulation in $\beta$, whereas CE trials yielded a slight downregulation, suggesting again that vision is critical for governing cortical activation – most significantly in high-frequency ($\gamma$ and $\beta$) waveforms. However, in habituation, the OE $\beta$ band remained highly active, but dropped significantly ($P < 0.05$) in CE trials. This suggests again both the governing role of the $\beta$ waveform in visually-aided cortical postural control, along with its diminishing utility without visual feedback.

Finally, the reported linearity results indicate some critical remarks regarding the degree of dependence between each paired variable; however, it is first important to define our interpretation of Pearson's correlation in this regard, wherein the greater the value of $R$, the lower the degree of inter-bandwidth independence and the more any observed intra-bandwidth significance may be due to natural spectral power variance. In adaptation, OE trials exhibited almost no linear correlation, with the exception of $\gamma_{high}$, whereas CE trials exhibited significant ($R > 0.7$) linear dependence in all bandwidths except $\alpha$ and $\gamma_{low}$. This distinction in comparing visual employment was antithetical in habituation trials, wherein OE linear dependence was significant ($R > 0.7$) in $\Delta$ and $\theta$ bands and strongly significant ($R > 0.9$) in both $\gamma_{low}$ and $\gamma_{high}$ bands, whereas CE trials yielded no linear correlation. These results altogether suggest disparate levels of dependence upon cortical activation when comparing OE and CE trials; in adaptation, the use of vision resulted in stronger inter-bandwidth independence, whereas in habituation, this independence was stronger without vision.

### 3.2. Functional Connectivity Analysis

Here, we explored two metrics for assessing functional connectivity during the three experimental phases: brain network segregation (using clustering coefficients, $CC$) and network integration (using betweenness centralities, $C$). The Jonckheere-Terpstra test was used to evaluate these network GTA metrics as a function of the level of control (from $BL$ to $Q1$ through $Q4$), with statistical significance determined by z-score equivalence (with $P$-values of node-wise comparisons corrected analogously using the BH false-discovery-rate method and $P < 0.05$ the threshold for significance). After implementing this methodology, only $\theta$-band segregation and $\alpha$-band network integration yielded statistical significance.

The top row in **Figure 4** shows CE network segregation cortical maps, denoting regions with a significant increase in the $\theta$ band $CC$ ($P < 0.05, corrected$) – a phenomenon mainly observed in central regions (left precentral, left postcentral, right paracentral) and parietal regions (left and right superioparietal). In addition, network integration maps illustrate a significant decrease in $\alpha$-band network integration ($P < 0.05, corrected$) from the experiment $BL$ to stimulation phases $Q1$ through $Q4$. These results mainly implicate cortical regions associated to the Default Mode Network (DMN: left anterior cingulate and left posterior cingulate), in addition to the cuneus and the left superiofrontal regions.

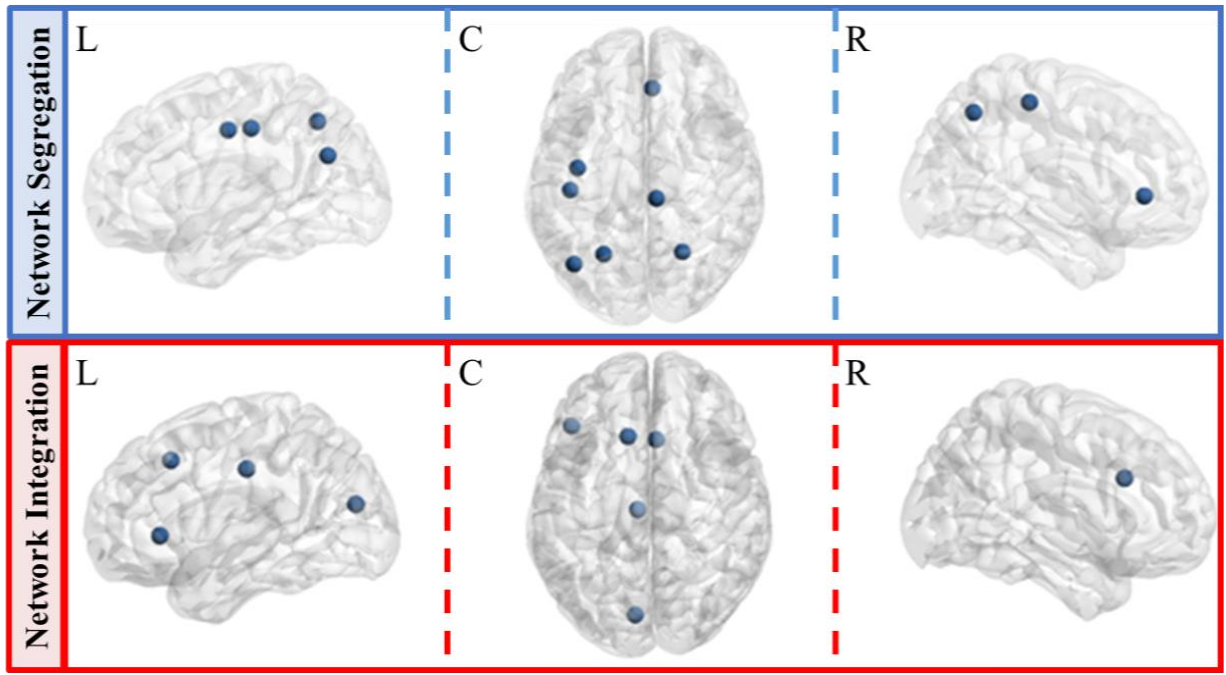

**Figure 4**: 3D representation of cortical brain regions, showing network segregation differences from $BL$ to $Q1$ to $Q4$ conditions within the $\theta$ band (top row) and network integration differences from $BL$, $Q1$ to $Q4$ conditions within the $\alpha$ band.

These results altogether indicate that cortical recruitment in postural control may be associated with decreased global functional connectivity at the $\theta$ band and decreased local functional connectivity at the $\alpha$ band. Before discussing the specific relevance of these findings, it is important to highlight the primary novelty of this type of investigation; this is the first reported employment of EEG source connectivity assessment to specifically characterize functional networks involved in postural control. While this notion makes it challenging to discuss the ultimate significance of these results, emerging literature does cite the utility of this method for investigating cognition [49] as well functional network changes in brain disorders [50], with correlative remarks to enable our discussion of this method in-context.

Our findings here suggest that functional brain networks are reshaped (i.e. undergo characteristic network modification) from a normative upright stance through perturbation in $Q1$ and $Q4$. This was most keenly expressed by the reported centro-parietal network's global characteristic shift in the $\theta$ waveform; recent investigation into scalp-level EEG analysis from activity in this centro-parietal network has reported similar changes from $BL$ to task failure of a suprapostural motor task with increasing postural destabilization [51]. The implication of parietal regions during postural control, in particular, was also recently reported using functional near-infrared spectroscopy [52]. Furthermore, frequency-based EEG source localization has implicated the $\theta$ band for its critical role in sensorimotor control and sensory information processing [53]. Our reported increase in local $\theta$ connectivity may therefore reflect activity occurring across the centro-parietal network that signifies either the planning of corrective steps and/or precise cognitive analyses of potential biomechanical consequences from incorrect or nonexistent postural correction.

Finally, the decrease in the $\alpha$ band network, located mainly in the DMN, may reflect an inhibition phenomenon (well-known for the $\alpha$ band), wherein the key role of error detection within the cingulate cortex was repressed – likely due to habituation [54, 55]. In addition, it was widely reported that the $\alpha$ network is highly involved in attentional processes by enabling the inhibition of irrelevant networks, influencing local/global signal processing, facilitating widespread information exchange, and enhancing perception [56].

## 4. Conclusions

Human postural control is a complex physiological task, reliant upon adaptive feedforward and feedback adjustments in the CNS, integrated with input from visual, vestibular, and somatosensory systems. While the roles of the CNS and subcortical structures in postural control is well-documented, there is still much debate in literature on the specific importance of the cerebral cortex and its relationship with vison. This was the motivation for the present work, wherein we describe the use of 64-channel EEG to investigate power spectral changes and cortical dynamics during the vibratory proprioceptive stimulation of calf muscles, measured with eyes open (OE) and eyes closed (CE).

The results presented here illustrate several principle conclusions. Firstly, power spectra variation analysis showed waveform-dependent activation within cortical regions specific to postural adaptation ($Q1 - BL$) and habituation ($Q4 - Q1$). In particular, the reported $\gamma$ band segmentation extends existing observations in literature that implicate its involvement in balance maintenance; we have shown here that instability engaged $\gamma_{low}$ and $\gamma_{high}$ waveforms analogously in OE and CE postural habituation, but the two bands behaved exactly the opposite in adaptation depending on the use of vision. Furthermore, generalized spectral variation resulted in significant spectral power shifts (from lower to higher frequencies) in CE adaptation trials, while generalized cortical activity decreased significantly in CE habituation trials. OE trials showed the opposite phenomenon, with both adaptation and habituation yielding increases in spectral power from $BL$ to $Q1$ and $Q1$ to $Q4$, respectively – albeit with fewer significant channels in each waveform. These results were further compared via pairwise variation, which illustrated the significant nature of our OE and CE observations; however, power spectra linearity results suggest that some EEG waveforms may not be strictly independent across trial conditions.

We have additionally reported novel results that reveal potential cortical networks involved in postural control using EEG source-space brain network mapping. Our reported increase in local $\theta$ connectivity may signify either the planning of corrective steps and/or precise cognitive analyses of falling consequences. In addition, $\alpha$ band network integration results may reflect an inhibition of error detection within the cingulate cortex, likely due to habituation. Our findings here suggest that functional brain networks undergo characteristic network modification; to our knowledge, these findings present the first reported characterization of both local and global brain network reshaping during postural control – a notion which necessitates future method optimization and the use of more subjects to support these results.

Altogether, the present findings suggest the critical notion that, following both acute and prolonged proprioceptive perturbation, cortical activity is upregulated with the availability of visual feedback, while conversely downregulated without vision. However, conclusions on inter-bandwidth independence would be strengthened by the use of more subjects, and implicated cortical regions would be further strengthened by the use of higher spatial resolution EEG systems. Furthermore, it should be made clear that the reported OE condition is a necessary but insufficient condition in itself for directly relating these findings to the availability of visual input. In addition, future investigation would be optimized by correlation with additional metrics for balance performance, such as posturography or EMG to gauge postural sway. Results from these included metrics may indeed help to clarify the feasibility and relevance of establishing the link between cortical response and body sway analysis. Nonetheless, the present study illustrates novel EEG spectral comparisons that provide evidence for the importance of visual referencing in postural control, and we further extend extant research by the introduction of adaptive cortical response analysis. These results invoke a novel emphasis on the importance of studying the role of the cortex in maintaining upright posture, providing further insight on challenging questions from a range of fields, from neurorehabilitation to the development of assistive tools for ambulation.

**Acknowledgments**

This research was supported jointly by the Institute for Biomedical and Neural Engineering at the University of Reykjavík, the Department of Anatomy at the University of Iceland, and the Icelandic National Hospital (Landspítali Scientific Fund, PI: Paolo Gargiulo) with additional funding support from the Rannís Icelandic Research Fund (Rannsóknasjodur, PI: Paolo Gargiulo). We would likewise like to acknowledge our international collaborators at the University of Rennes, the School of Life and Health Sciences at Aston University, and the Department of Electrical Engineering and Information Technology at the University of Naples, Federico II.

**Disclosures**

The authors declare that there is no conflict of interest regarding the publication of this paper.

**Ethics Approval and Consent to Participate**

Ethical approval for Icelandic subject data acquisition, as reported herein in this study, was obtained by the Icelandic Science and Ethics Committee (RU Code of Ethics, cf. Paragraph 3 in Article 2 of the Higher Education Institution Act no. 63/2006). All information regarding subject identity is confidential in accordance with the "Act on the Protection of Privacy as regards the Processing of Personal Data", No. 77/2000. All subjects have signed an informed consent agreement regarding their participation and the anonymous publication of acquired data; consent forms were prepared according to guidelines set by the Icelandic Bioethics Committee (Vísindasiðanefnd, Hafnarhæusið, Tryggvagata 17, 101

Reykjavík Iceland) with approval reference ID 13-127-S1. No specific acquired data will be made available to external entities, and any and all electronic data that may be distributed between collaborating individuals is entirely anonymous.